\documentclass[conference]{IEEEtran}
\IEEEoverridecommandlockouts
\usepackage{cite}
\usepackage{amsmath,amssymb,amsfonts}
\usepackage{amsthm}
\usepackage{algorithmic}
\usepackage{graphicx}
\usepackage{textcomp}
\usepackage{xcolor}
\usepackage[normalem]{ulem} 
\usepackage{arydshln} 
\usepackage{verbatim}

\newcommand{\gh}[1]{\textcolor{magenta}{#1}}

\def\BibTeX{{\rm B\kern-.05em{\sc i\kern-.025em b}\kern-.08em
    T\kern-.1667em\lower.7ex\hbox{E}\kern-.125emX}}
\begin{document}

\title{Solos: A Dataset for Audio-Visual Music Analysis \\
\thanks{
This work has received funding from the MICINN/FEDER UE project with reference PGC2018-098625-B-I00,  H2020-MSCA-RISE-2017 project with reference 777826 NoMADS, ERC Innovation Programme (grant 770376, TROMPA),
Spanish Ministry of Economy and Competitiveness under the María de Maeztu Units of Excellence Program (MDM-2015-0502)  and the Social European Funds. We also thank Nvidia for the donation of GPUs.}
}

\author{
    \IEEEauthorblockN{Juan F. Montesinos, Olga Slizovskaia, Gloria Haro}
    \IEEEauthorblockA{Department of Information and Communications Technologies \\
    Universitat Pompeu Fabra, Barcelona, Spain \\ \{juanfelipe.montesinos, olga.slizovskaia, gloria.haro\}@upf.edu}}

\maketitle
\begin{abstract}

In this paper, we present a new dataset of music performance videos which can be used for training machine learning methods for multiple tasks such as audio-visual blind source separation and localization,  cross-modal correspondences, cross-modal generation and, in general, any audio-visual self-supervised task. These videos, gathered from YouTube, consist of solo musical performances of 13 different instruments. Compared to previously proposed audio-visual datasets, Solos is cleaner since a big amount of its recordings are auditions and manually checked recordings, ensuring there is no background noise nor effects added in the video post-processing. Besides, it is, up to the best of our knowledge, the only dataset that contains the whole set of instruments present in the URMP\cite{URPM} dataset, a high-quality dataset of 
44  audio-visual recordings of multi-instrument classical
music pieces with individual audio tracks. URMP was intented to be used for source separation, thus, we evaluate the performance on the URMP dataset of two different source-separation models trained on Solos. The dataset is publicly available at https://juanfmontesinos.github.io/Solos/
\end{abstract}


\section{Introduction}
There is a growing interest in multimodal techniques for solving  Music Information Retrieval (MIR) problems. 
Music performances have a highly multimodal content and the different modalities involved are highly correlated: sounds are emitted by the motion of the player performing and in chamber music performances the scores constitute an additional encoding that may be as well leveraged for the automatic analysis of music \cite{li2017see}.
 
A fundamental problem in music analysis and in general in audio processing is Blind Source Separation (BSS). BSS consists in, given a mixture of signals, recovering the individual signals the mixture is conformed by. Mathematically, a mixture of sounds can be expressed as the sum of individual sources: $s_m = \sum s_i$. Thus, the BSS problem consists in recovering each $s_i$ for a given $s_m$. In speech, it is also known as the Cocktail Party problem, which refers to the task
of recognizing an individual speech in noisy social environments \cite{cherry1953some}. 
Single-channel source separation problem can be approached from an audio-only perspective using techniques
such as independent component analysis (ICA) \cite{hyvarinen2000independent}, sparse decomposition \cite{zibulevsky2001blind}, nonnegative
matrix factorization (NMF) \cite{virtanen2007monaural}, computational auditory scene analysis (CASA) \cite{ellis1996prediction}, probabilistic
latent component analysis (PLCA) \cite{smaragdis2006probabilistic} or deep learning techniques (e.g. \cite{chandna2017monoaural, stoller2018wave }).

On the other side, by
visually inspecting the scene we may extract information about the number of sound sources, their
type, spatio-temporal location and also motion, which naturally relates to the emitted sound. Besides, it is possible to carry out self-supervised tasks in which one modality supervises the other one. This entails another research field, the cross-modal correspondence (CMC). We can find pioneering works for both problems BSS and CMC.  \cite{hershey2000audio, kidron2005pixels} make use of audio-visual data for sound localization and \cite{darrell2000audio}, \cite{sodoyer2002separation},\cite{rivet}  for speech separation. In the context of music, visual information has also proven to help model-based methods both in source separation \cite{li2017audiovisual, parekh2017guiding} and localization \cite{li2017see}. 
With the flourishing of deep learning techniques many recent works exploit both, audio and video content, to perform music source separation \cite{gao2019co, SoM, xu2019recursive}, source association \cite{li2019online}, localization \cite{arandjelovic2018objects} or both \cite{SoP}. Some CMC works explore features generated from synchronization \cite{owenssync,kobar} and prove these features are reusable for source separation. These works use networks that have been trained in a self-supervised way using pairs of corresponding/non-corresponding audio-visual signals for localization purposes \cite{arandjelovic2018objects} or the mix-and-separate approach for source separation \cite{SoP, gao2019co, SoM, xu2019recursive}. Despite deep learning made possible to solve classical problems in a different way, it also contributed to create new research fields like cross-modal generation, in which the main aim is to generate video from audio\cite{speech2face,rochestergen} or viceversa \cite{visual2sound}. More recent works related to human motion make use of skeleton as an inner representation of the body which can be further converted into video \cite{audio2body,conversationalgestures} which shows the potential of skeletons. 
The main contribution of this paper is Solos, a new dataset of musical performance recordings of soloists that can be used to train deep neural networks for any of the aforementioned fields. Compared to a similar dataset of musical instruments presented in \cite{SoP} and its extended version \cite{Musices}, our dataset does contain the same type of chamber orchestra instruments present in the URMP dataset. Solos is a dataset of 755 real-world recordings gathered from YouTube which provides several features missing in the aforementioned datasets:  skeletons and high quality timestamps. Source localization is usually indirectly learned by networks. Thus, providing a practical localization ground-truth is not straightforward. Nevertheless, networks often point to the player hands as if they were the sound source. We expect hands localization can help to provide additional cues to improve audio-visual BSS or can be used as source ground-truth localization. In order to show the benefits of using Solos we trained some popular BSS architectures and compare their results.

\section{Related Work}

The University of Rochester Multi-Modal Music Performance Dataset (URMP) \cite{URPM} is a dataset with 44 multi-instrument video recordings of classical music pieces. Each instrument present in a piece was recorded separately, both with video and high-quality audio with a stand-alone microphone, in order to have ground-truth individual tracks. Although playing separately, the instruments were coordinated by using a conducting video with a pianist playing in order to set the common timing for the different players. After synchronization, the audio of the individual videos was replaced by the high-quality audio of the microphone and then different recordings were assembled to create the mixture: the individual high-quality audio recordings were added up to create the audio mixture and the visual content was composited in a single video with a common background where all players were arranged at the same level from left to right. For each piece, the dataset provides the musical score in MIDI format, the high-quality individual instrument audio recordings and the videos of the assembled pieces. The instruments present in the dataset, shown in Figure \ref{fig: categories}, are common instruments in chamber orchestras. In spite of all its good characteristics, it is a small dataset and thus not appropriate for training  deep learning architectures.

Two other datasets of audio-visual recordings of musical instruments performances have been presented recently: 
Music \cite{SoP} and MusicES \cite{Musices}. Music consists of 536 recordings of solos and 149 videos of duets across 11 categories: accordion, acoustic guitar, cello, clarinet, erhu, flute, saxophone, trumpet, tuba, violin and xylophone. This dataset was gathered by querying YouTube. MusicES \cite{Musices} is an extension of MUSIC to around the triple of its original size with approximately  1475 recordings but spread in 9 categories instead: accordion, guitar, cello, flute, saxophone, trumpet, tuba, violin and xylophone.  There are 7  common categories in MUSIC and Solos: violin, cello, flute, clarinet,  saxophone, trumpet and tuba. The common categories between MusicES and Solos are 6 (the former ones except clarinet). Solos and MusicES are complementary.  There is only an small intersection of 5\% between both, which means both datasets can be combined into a bigger one.

We can find in the literature several examples which show the utility of audio-visual datasets. \textit{The Sound of Pixels} \cite{SoP} performs audio source separation generating audio spectral components which are further smartly selected by using visual features coming from the video stream to obtain separated sources. This idea was further extended in \cite{xu2019recursive} in order to separate the different sounds present in the mixture in a recursive way. At each stage, the system separates the most salient source from the ones remaining in the mixture.  \textit{The Sound of Motions} \cite{SoM} uses dense trajectories obtained from optical flow to condition audio source separation, being able even to separate same-instrument mixtures. Visual conditioning is also used in \cite{gao2019co} to separate different instruments; during training, a classification loss is used on the separated sounds to enforce object consistency and a co-separation loss forces the estimated individual sounds to produce the original mixtures once reassembled. 
In 
\cite{parekh2017guiding}, the authors developed an energy-based method which minimizes a \textit{Non-Negative Matrix Factorization} term with an activation matrix which is forced to be aligned to a matrix containing per-source motion information. This motion matrix  
contains the average magnitude velocities of the clustered motion trajectories in each player bounding box.

Recent works show the rising use of skeletons in audiovisual tasks. In  \textit{Audio to body dynamics} \cite{audio2body} authors show it is possible to predict skeletons reproducing the movements of players playing instruments such as piano or violin. 
Skeletons have proven to be useful for establishing audio-visual correspondences, such as body or finger motion with note onsets or pitch fluctuations, in chamber music performances  \cite{li2019online}. A recent work
\cite{music2gesture} tackles the source separation problem in a 
 similar to \textit{Sound of Motions} \cite{SoM} but  replacing the dense trajectories by skeleton information. 

\section{Dataset}
\begin{figure}[t]
\includegraphics[width=0.5\textwidth]{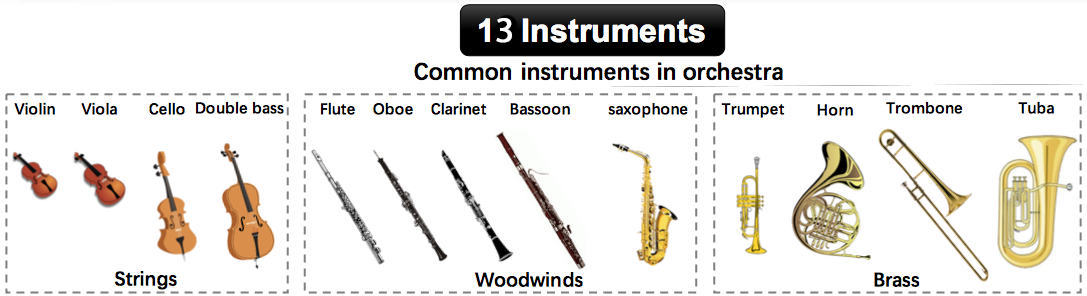}
\caption{Solos and URMP instrument categories. Image adapted from \cite{URPM}.}
\label{fig: categories}
\end{figure}

Solos\footnote{\label{github}Dataset available at https://juanfmontesinos.github.io/Solos/}
was designed to have the same categories as the URMP \cite{URPM} dataset, so that URMP can be used as testing dataset in a real-world scenario. This way we aim to establish a standard way of evaluating source separation algorithms' performance avoiding the use of mix-and-separate  in testing.
Solos consists of 755 recordings distributed amongst 13 categories as shown in Figure \ref{fig: categories}, with an average amount of 58 recordings per category and an average duration of 5:16 min. It is interesting to highlight that, for 8 out of 13 categories, the median of resolution is HD, despite being a YouTube-gathered dataset. Per-category statistics can be found in Table \ref{tab:solos_stats}. These recordings were gathered by querying YouTube using the tags solo and auditions in several languages such as English, Spanish, French, Italian, Chinese or Russian. 

\begin{table}[]
\centering
\begin{tabular}{lccc}
\multicolumn{1}{l}{Category} & \# Recordings & Mean duration & Median resolution \\ \hline
Violin                       & 66            & 6:16          & 1080x720          \\
Viola                        & 55            & 5:31          & 1280x720          \\
Cello                        & 134           & 7:21          & 640x480           \\
DoubleBass                   & 58            & 8:53          & 1280x720          \\ \hdashline

Flute                        & 48            & 4:00          & 640x360           \\
Oboe                         & 53            & 5:45          & 1280x720          \\
Clarinet                     & 49            & 3:23          & 640x360           \\
Bassoon                      & 56            & 5:08          & 1280x720          \\
Saxophone                    & 45            & 2:42          & 1280x720          \\ \hdashline

Trumpet                      & 50            & 1:14          & 640x360           \\
Horn                         & 50            & 5:11          & 1280x720          \\
Trombone                     & 50            & 5:03          & 1280x720          \\ 
Tuba                         & 41            & 2:49          & 640x360           \\ \hdashline

{\bf TOTAL}                  & {\bf 755}     & {\bf 5:16}    & {\bf 854x480}          
\end{tabular}
\caption{Statistics of Solos Dataset}
\label{tab:solos_stats}
\end{table}
\subsection{OpenPose Skeletons}
Solos is not only a set of recordings. Apart from the videos identificators We also provide: i) body and hand  skeletons estimated by OpenPose \cite{openpose} in each frame of each recording and ii) timestamps indicating useful parts. OpenPose is a system capable to predict body skeleton and hands skeletons making use of two different neural networks. To do so, they predict a confidence map of the belief that a specific body part may be located at any given pixel as well as part affinity fields which encode the degree of association between different body parts. Finally, it predicts 2D skeletons and per-joint confidence via greedy inference. In practice, the body skeleton is estimated with a first network. Then, the position  of the wrists in the body skeleton are used to estimate the position of both hands. A second neural network obtains the skeleton of each hand independently. Note that since each body part is estimated independently, OpenPose makes no assumptions about the limbs to find. It just calculates the most likely skeleton given confidence maps and part affinity fields. The whole process is carried out frame-wise. This leads to a small flickering and mispredictions  between frames. 

\subsection{Timestamps estimation and skeleton refinement}
Video streams are re-sampled to 25 FPS keeping the audio stream intact. An iterative process returns stamps for which there are at least N frames with a detected hand and no more than M consecutive mispredictions. In practice we use N=150 and M=5, thus, a minimum of 6 seconds of video with at most 5 consecutive frames with hand mispredictions. At this point, we have segments of video in which there are hands detected. To refine these results we further applied an energy-based silence detector which allows to discard those segments in which the instrument is not being played, e.g., transitions, music sheet changes, etcetera. Besides, we perform a linear interpolation of the mispredicted keypoints in a relative base of coordinates. 
Directly interpolating the absolute coordinates would lead to deformations of the skeleton and inaccuracies.  Since skeletons are tree-like graphs  it is possible to interpolate the relative coordinates of each joint (node in the graph) with respect to its parent node. Then, the absolute coordinates of the joint are recovered with the sum of the absolute coordinates of its parent and the estimated relative coordinates with respect to the parent. Let us denote by $J_i
^t$ the relative coordinates of the $i$th joint with respect to its parent at time $t$. On the other hand, $\hat{J}^t_i$ denotes the estimated value of $J_i^t$ when the $i$th joint is mispredicted. $\hat{J}^t_i$ can be linearly interpolated using the relative coordinates of the closest $i$th detected joint before time $t$ (i.e $J^{t^-}_i$ where $t^- < t$), and analogously with the closest $i$th detected joint after time $t$ (i.e $J^{t^+}_i$ where $t < t
^+$). For example, given the following sequence of detected and misdetected coordinates (that need to be estimated), $J$ and $\hat{J}$ respectively:
\begin{equation*}
\label{eq:set}
\{J_i^{t-n},\hat{J}_i^{t-n+1},...,\hat{J}_i^{t},...,\hat{J}_i^{t+m-1},J_i^{t+m}\}
\end{equation*}
then, the interpolation at time $t$ can be calculated as:
\begin{equation}
\label{eq:interp}
\hat{J}^t_i=\frac{m \, J_i^{t-n}  + n \, J_i^{t+m}}{m+n}.
\end{equation}
OpenPose maps mispredicted joints to the origin of coordinates. We empirically found that such a big jump in the position of a joint induces noise. Using interpolated coordinates helps to address this problem. 
\vspace{6mm}

\section{Experiments}
In order to show the suitability of Solos, we have focused in the blind source separation problem and have trained  \textit{The Sound of Pixels} (SoP) \cite{SoP} and the Multi-head U-Net (MHU-Net) \cite{multihead} models on the new dataset. We have carried out four experiments: i) we have evaluated the SoP pre-trained model  provided by the authors; ii) we have trained SoP from  scratch; iii) we have fine-tuned SoP on Solos starting from the weights of the pre-trained model on MUSIC and iv) we have trained the Multi-head U-Net from scratch. 
MHU-Net has been trained to separate mixtures with the number of sources varied from two to seven following a curriculum learning procedure as it improves the results. SoP has been trained according to the optimal strategy described in \cite{SoP}.
\vspace{4mm}
\begin{figure*}[t]
\centering
\includegraphics[width=\textwidth]{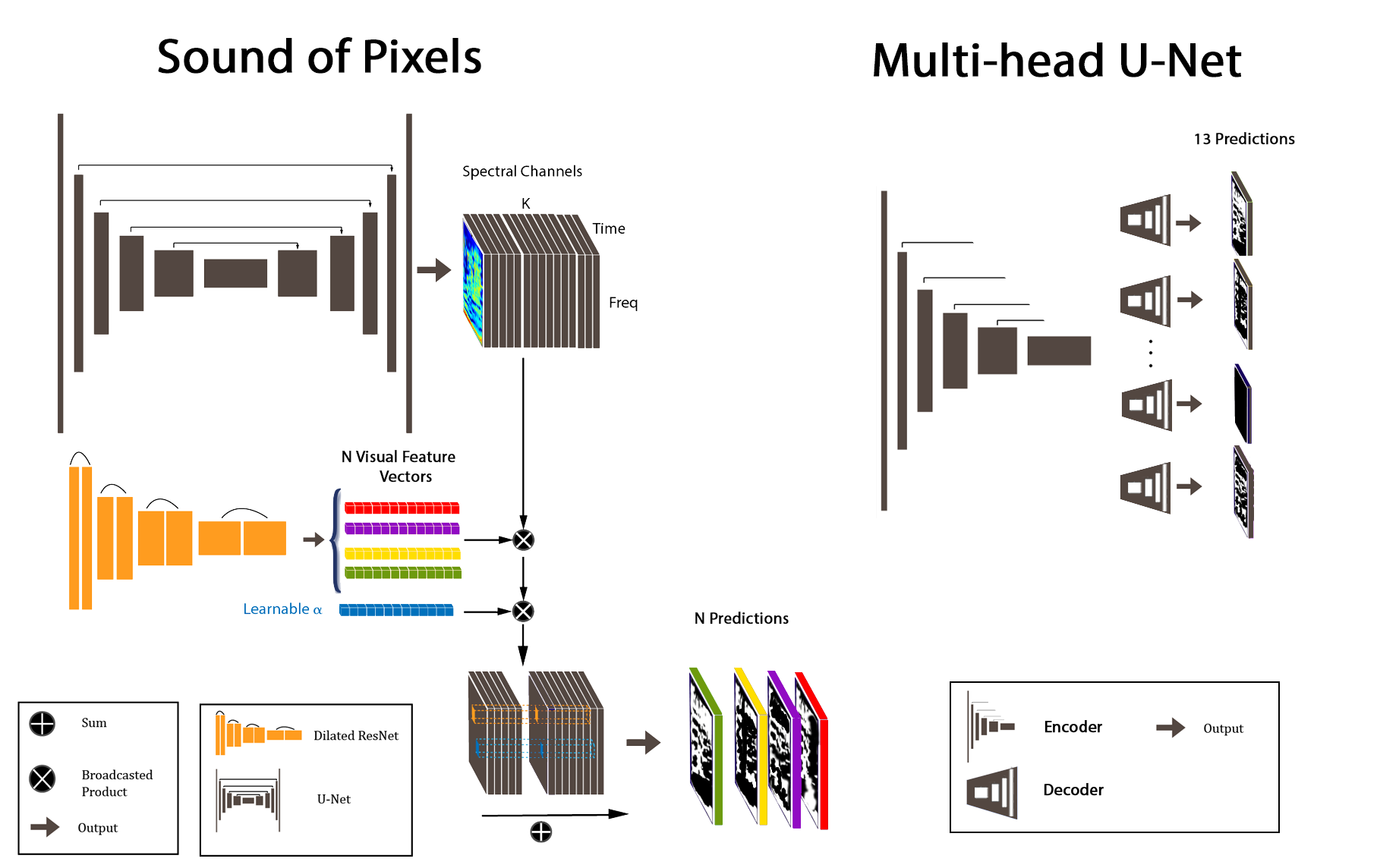}
\caption{Considered architectures. Left, Sound of Pixels: The network takes as input a mixture spectrogram and returns a binary mask given the visual feature vector of the desired source. Right, Multi-Head U-Net: It takes as input a mixture spectrogram and returns 13 ratio masks, one per decoder. }
\label{fig: models}
\end{figure*}

Evaluation is performed on the URMP dataset \cite{URPM} using the real mixtures they provide. URMP tracks are sequentially split in 6s-duration segments. Metrics are obtained from all the resulting splits.

\subsection{Architectures and training details}
We have chosen \textit{The Sound of Pixels} as baseline since its weights are publicly available and the network is trained in a straight-forward way. SoP is composed of three main sub-networks: A dilated ResNet\cite{DResnet} as  video-analysis network, a U-Net\cite{unet} as  audio-processing network and an audio synthesizer network.  We also compare its results against a Multi-head U-Net \cite{multihead}.
\vspace{4mm}

U-Net  \cite{ronneberger2015u} is an encoder-decoder architecture with skip connections in between. Skip connections help to recover the original spatial structure. 
MHU-Net is a step forward as it consist of as many decoders as possible sources. Each decoder is specialized in a single source, thus improving performance. 

\textit{The Sound of Pixels}\cite{SoP} does not follow the original U-Net architecture proposed for biomedical imaging, but the U-Net described at \cite{unet}, which was tuned for singing voice separation. Instead of having two convolutions per block followed by max-pooling, they use a single convolution with a bigger kernel and striding. The original work  proposes a central block with learnable parameters whereas the central block is a static latent space in \textit{SoP}. U-Net has been widely used as backbone of several architectures for tasks such us image generation \cite{unet-img-synt}, noise suppression and super-resolution \cite{mao2016image}, image-to-image translation\cite{pix2pix}, image segmentation \cite{ronneberger2015u} or audio source separation \cite{unet}.
SoP U-Net consists of 7 blocks with 32, 64, 128, 256, 512, 512 and 512 channels respectively (6 blocks for the MHU-Net). The latent space can be considered as the last output of the encoder. 
Dilated ResNet is a ResNet-like architecture which makes use of dilated convolutions to keep the receptive field while increasing the resulting spatial resolution. 
The output of the U-Net is a set of 32 spectral components (channels) which are the same size than the input spectrogram, in case of SoP, and a single source per decoder in case of MHU-Net. Given a representative frame, visual features are obtained using the Dilated ResNet. These visual features are nothing but a vector of 32 elements (which corresponds to the number of output channels of U-Net) which are used to select proper spectral components. This selection is performed by the audio analysis network which consist of 32 learnable parameters, $\alpha_k$,  plus a bias, $\beta$. This operation can be mathematically described as follows:

\begin{equation*}
    \beta+\sum_{k=1}^{32} \alpha_k v_{f_k}S_k(t,f),
\end{equation*}
where $S_k(t,f)$ is the $k$-th predicted spectral component at time-frequency bin $(t,f)$.

Figure \ref{fig: models} illustrates the SoP configuration. It is interesting to highlight that making the visual network to  select the spectral components forces it to indirectly learn instrument localization, which can be inferred via activation maps.
\vspace{4mm}

On one hand, MHU-Net has been trained using a curriculum learning strategy that consists of a gradual increment on the amount of sources present in the mixture from two to four. When the loss stays on a plateau for more than 160,000 iterations, the amount of sources is increased by one. We have used mean-square error loss, ADAM  optimizer \cite{Adam}, an initial learning rate (LR)  of $10^{-4}$, weight decay of $10^{-5}$  and dropout of $0.2$ in the decoder. We have also reduced the LR by a half if the loss stays on a plateau for more than 400,000 iterations. 

On the other hand, SoP has been trained using a LR of $10^{-3}$ for the U-Net and a LR of $10^{-4}$ for the Dilated ResNet as it was pre-trained on ImageNet.
We have applied a weight on the gradients  based on the magnitude of the mixture spectrogram so that time-frequency points of the predicted source/s contribute to the loss according to the energy of the analogous time-frequency points in the mixture spectrogram.  This reduces overfitting since, given a source, 
a time-frequency bin with a low value may be assigned either to one or zero in the ground-truth mask depending on the recorded noise and such weights help reduce its impact on the training. We used different training strategies for SoP and MHU-Net as the optimal training for SoP harms the performance of the MHU-Net. 

Ground-truth mask calculation for both SoP and MHU-Net are  described in Eq. \eqref{eq:mask} and Eq. \eqref{eq:mask_ratio}, Sec. \ref{sec: GT}.

\subsection{Data pre-processing}
In order to train the aforementioned architectures, audio is re-sampled to 11025 Hz and 16 bit. Samples fed into the network are 6s duration. We use Short-time Fourier Transform (STFT) to obtain time-frequency representations of waveforms. Following \cite{SoP}, STFT is computed using Hanning window of length 1022 and hop length 256 so that we obtain a spectrogram of size 512$\times$256 for a 6s sample. Later on, we apply a log re-scale on the frequency axis expanding lower frequencies and compressing higher ones. Lastly, we convert  magnitude spectrograms into dB w.r.t. the minimum value of each spectrogram and normalize between -1 and 1. 
\vspace{4mm}
\subsection{Ground-truth mask} \label{sec: GT}
Before introducing ground-truth mask computations we would like to point out some considerations. Standard  floating-point audio format imposes a waveform to be bounded between -1 and 1.  At the time of creating artificial mixtures resulting waveforms may be out of these bounds. This can help neural networks to find shortcuts to overfit. To avoid this behaviour spectrograms are clamped according to the equivalent bounds in the time-frequency domain.
\vspace{4mm}

The Discrete Short-time Fourier Transform  can be computed as described in \cite{STFT}:
\begin{equation*}
    S[t,f] = \sum_{k=0}^{L-1}s[k]w[k-t]e^{-j\pi fk/L}.
\end{equation*}
Since $s[k] \in [-1,1]$ it can be easily shown that:
$$
    |S[t,f]|  \leq \sum_{k=0}^{L-1}|w[k-t]|,
$$
i.e., that the  magnitude STFT of an audio signal bounded between [-1,1] is bounded between $[0,\sum |w[k]|]$. Thus, 
given  the STFT of N waveforms,  the spectogram of a mixture of sounds is defined the following way:
\begin{equation*}
\label{eq: mix}
     S_{mix}(t,f) = \min\left(\sum_{n=1}^N S_n(t,f),\sum |w[k]|\right),
\end{equation*}
which is equivalent to:
\begin{equation*}
\label{eq: mix_time}
     S_{mix}(t,f) = STFT \left\{\min \left(1,\max\left(\sum_{n=1}^N s_n(t),-1\right)\right)\right\}.
\end{equation*}
For training Sound of Pixels  we have used complementary binary masks as ground-truth masks, defined as:

\begin{equation}
\label{eq:mask}
   M_n(t,f) =
    \begin{cases}
      1, & \text{if}\  |S_n(t,f)|\geq |S_{m}(t,f)| \, \, \, \forall m \in \{1, ..., N\},\\
      0, & \text{otherwise}.
    \end{cases}
\end{equation}
The Multi-head U-Net has been trained with complementary ratio masks, defined as:
\begin{equation}
\label{eq:mask_ratio}
   M_n(t,f) = \frac{|S_n(t,f)|}{ |S_{mix}(t,f)|}.
\end{equation}

\subsection{Results}
Benchmark results for Source to Distortion Ratio (SDR), Source to Interferences Ratio (SIR), Sources to Artifacts Ratio (SAR) proposed in \cite{sdr} are shown in Table \ref{tab:benchmark} in terms of mean and standard deviation.
As it can be observed, Sound of Pixels evaluated using its original weights performs the worst. One possible reason for that could be the absence of some of the URMP categories on the  MUSIC dataset. If we train the network from scratch on Solos, results improve by almost 1 dB.  However, it is possible to achieve an even better result fine-tuning the network, pre-trained with MUSIC, on Solos. We hypothesize that the improvement occurs as the network is exposed to much more training data. Moreover, the table results show how it is possible to reach higher performance by using more powerful architectures like MHU-Net.
\begin{table}[]
    \centering
    \begin{tabular}{c|c|c|c}
          & SDR $\uparrow$ & SIR $\uparrow$ & SAR $\uparrow$ \\ \hline
         SoP \cite{SoP} & $-3.76\pm4.00$ & $-1.45\pm4.68$ & $7.56\pm3.13$ \\
         SoP-Solos & $-2.98\pm5.07$ & $0.46\pm6.76$ & $6.37\pm2.94$ \\
         SoP-ft & $-2.57\pm4.99$ & $0.47\pm6.43$ & $6.89\pm2.48$ \\
         MHU-Net & $ -0.56\pm5.96 $ & $ 1.04\pm7.24 $ & $ 10.37\pm3.48 $ \\ 
    \end{tabular}
    \caption{Benchmark results (mean $\pm$ standard deviation). SoP:Sound of Pixels original weights, SoP-Solos: Sound of Pixels trained from scratch on Solos. SoP-ft: Sound of Pixels finetuned on Solos. MHU-Net: Multi-head U-Net with 13 decoders. }
    \label{tab:benchmark}
\end{table}

\section{Conclusions}
We have presented Solos, a new audio-visual dataset of music recordings of soloists, suitable for different self-supervised learning tasks such as source separation  using the mix-and-separate strategy, sound localization, cross-modal generation and finding audio-visual correspondences. There are 13 different instruments in the dataset; those are common instruments in chamber orchestras and the ones included in the University  of  Rochester  Multi-Modal  Music  Performance (URMP) dataset \cite{URPM}. The characteristics of URMP -- small dataset of real performances with ground truth individual stems -- make it a suitable dataset for testing purposes but to the best of our knowledge, to date there is no existing large-scale dataset with the same instruments as in URMP. Two different networks for audio-visual source separation based on the U-Net architecture have been trained in the new dataset and further evaluated in URMP, showing the impact of training on the same set of instruments as the test set. Moreover, Solos provides skeletons and timestamps to video intervals where hands are sufficiently visible. This information could be useful for training purposes and also for learning to solve the task of sound localization.

\bibliographystyle{IEEEtran}
\bibliography{IEEEabrv,mybib}

\begin{thebibliography}{10}
\providecommand{\url}[1]{#1}
\csname url@samestyle\endcsname
\providecommand{\newblock}{\relax}
\providecommand{\bibinfo}[2]{#2}
\providecommand{\BIBentrySTDinterwordspacing}{\spaceskip=0pt\relax}
\providecommand{\BIBentryALTinterwordstretchfactor}{4}
\providecommand{\BIBentryALTinterwordspacing}{\spaceskip=\fontdimen2\font plus
\BIBentryALTinterwordstretchfactor\fontdimen3\font minus
  \fontdimen4\font\relax}
\providecommand{\BIBforeignlanguage}[2]{{%
\expandafter\ifx\csname l@#1\endcsname\relax
\typeout{** WARNING: IEEEtran.bst: No hyphenation pattern has been}%
\typeout{** loaded for the language `#1'. Using the pattern for}%
\typeout{** the default language instead.}%
\else
\language=\csname l@#1\endcsname
\fi
#2}}
\providecommand{\BIBdecl}{\relax}
\BIBdecl

\bibitem{URPM}
B.~{Li}, X.~{Liu}, K.~{Dinesh}, Z.~{Duan}, and G.~{Sharma}, ``Creating a
  multitrack classical music performance dataset for multimodal music analysis:
  Challenges, insights, and applications,'' \emph{IEEE Transactions on
  Multimedia}, vol.~21, no.~2, pp. 522--535, Feb 2019.

\bibitem{li2017see}
B.~Li, K.~Dinesh, Z.~Duan, and G.~Sharma, ``See and listen: Score-informed
  association of sound tracks to players in chamber music performance videos,''
  in \emph{2017 IEEE International Conference on Acoustics, Speech and Signal
  Processing (ICASSP)}.\hskip 1em plus 0.5em minus 0.4em\relax IEEE, 2017, pp.
  2906--2910.

\bibitem{cherry1953some}
E.~C. Cherry, ``Some experiments on the recognition of speech, with one and
  with two ears,'' \emph{The Journal of the acoustical society of America},
  vol.~25, no.~5, pp. 975--979, 1953.

\bibitem{hyvarinen2000independent}
A.~Hyv{\"a}rinen and E.~Oja, ``Independent component analysis: algorithms and
  applications,'' \emph{Neural networks}, vol.~13, no. 4-5, pp. 411--430, 2000.

\bibitem{zibulevsky2001blind}
M.~Zibulevsky and B.~A. Pearlmutter, ``Blind source separation by sparse
  decomposition in a signal dictionary,'' \emph{Neural computation}, vol.~13,
  no.~4, pp. 863--882, 2001.

\bibitem{virtanen2007monaural}
T.~Virtanen, ``Monaural sound source separation by nonnegative matrix
  factorization with temporal continuity and sparseness criteria,'' \emph{IEEE
  transactions on audio, speech, and language processing}, vol.~15, no.~3, pp.
  1066--1074, 2007.

\bibitem{ellis1996prediction}
D.~P.~W. Ellis, ``Prediction-driven computational auditory scene analysis,''
  Ph.D. dissertation, Massachusetts Institute of Technology, 1996.

\bibitem{smaragdis2006probabilistic}
P.~Smaragdis, B.~Raj, and M.~Shashanka, ``A probabilistic latent variable model
  for acoustic modeling,'' \emph{Advances in models for acoustic processing,
  NIPS}, vol. 148, pp. 8--1, 2006.

\bibitem{chandna2017monoaural}
P.~Chandna, M.~Miron, J.~Janer, and E.~G{\'o}mez, ``Monoaural audio source
  separation using deep convolutional neural networks,'' in \emph{International
  Conference on Latent Variable Analysis and Signal Separation}, 2017, pp.
  258--266.

\bibitem{stoller2018wave}
D.~Stoller, S.~Ewert, and S.~Dixon, ``Wave-u-net: A multi-scale neural network
  for end-to-end audio source separation,'' \emph{arXiv preprint
  arXiv:1806.03185}, 2018.

\bibitem{hershey2000audio}
J.~R. Hershey and J.~R. Movellan, ``Audio vision: Using audio-visual synchrony
  to locate sounds,'' in \emph{Advances in neural information processing
  systems}, 2000, pp. 813--819.

\bibitem{kidron2005pixels}
E.~Kidron, Y.~Y. Schechner, and M.~Elad, ``Pixels that sound,'' in
  \emph{Computer Vision and Pattern Recognition, 2005. CVPR 2005. IEEE Computer
  Society Conference on}, vol.~1, 2005, pp. 88--95.

\bibitem{darrell2000audio}
T.~Darrell, J.~W. Fisher, and P.~Viola, ``Audio-visual segmentation and “the
  cocktail party effect”,'' in \emph{Advances in Multimodal Interfaces—ICMI
  2000}, 2000, pp. 32--40.

\bibitem{sodoyer2002separation}
D.~Sodoyer, J.-L. Schwartz, L.~Girin, J.~Klinkisch, and C.~Jutten, ``Separation
  of audio-visual speech sources: a new approach exploiting the audio-visual
  coherence of speech stimuli,'' \emph{EURASIP Journal on Advances in Signal
  Processing}, vol. 2002, no.~11, p. 382823, 2002.

\bibitem{rivet}
B.~{Rivet}, L.~{Girin}, and C.~{Jutten}, ``Mixing audiovisual speech processing
  and blind source separation for the extraction of speech signals from
  convolutive mixtures,'' \emph{IEEE Transactions on Audio, Speech, and
  Language Processing}, vol.~15, no.~1, pp. 96--108, 2007.

\bibitem{li2017audiovisual}
B.~Li, C.~Xu, and Z.~Duan, ``Audiovisual source association for string
  ensembles through multi-modal vibrato analysis,'' \emph{Proc. Sound and Music
  Computing (SMC)}, 2017.

\bibitem{parekh2017guiding}
S.~Parekh, S.~Essid, A.~Ozerov, N.~Q. Duong, P.~P{\'e}rez, and G.~Richard,
  ``Guiding audio source separation by video object information,'' in
  \emph{Applications of Signal Processing to Audio and Acoustics (WASPAA), 2017
  IEEE Workshop on}, 2017, pp. 61--65.

\bibitem{gao2019co}
R.~Gao and K.~Grauman, ``Co-separating sounds of visual objects,'' in
  \emph{Proceedings of the IEEE International Conference on Computer Vision},
  2019, pp. 3879--3888.

\bibitem{SoM}
H.~Zhao, C.~Gan, W.-C. Ma, and A.~Torralba, ``The sound of motions,'' in
  \emph{Proceedings of the IEEE International Conference on Computer Vision},
  2019, pp. 1735--1744.

\bibitem{xu2019recursive}
X.~Xu, B.~Dai, and D.~Lin, ``Recursive visual sound separation using minus-plus
  net,'' in \emph{Proceedings of the IEEE International Conference on Computer
  Vision}, 2019, pp. 882--891.

\bibitem{li2019online}
B.~Li, K.~Dinesh, C.~Xu, G.~Sharma, and Z.~Duan, ``Online audio-visual source
  association for chamber music performances,'' \emph{Transactions of the
  International Society for Music Information Retrieval}, vol.~2, no.~1, 2019.

\bibitem{arandjelovic2018objects}
R.~Arandjelovi{\'c} and A.~Zisserman, ``Objects that sound,'' in
  \emph{Proceedings of the IEEE European Conference on Computer Vision}, 2018.

\bibitem{SoP}
H.~Zhao, C.~Gan, A.~Rouditchenko, C.~Vondrick, J.~McDermott, and A.~Torralba,
  ``The sound of pixels,'' in \emph{The European Conference on Computer Vision
  (ECCV)}, September 2018.

\bibitem{owenssync}
A.~Owens and A.~A. Efros, ``Audio-visual scene analysis with self-supervised
  multisensory features,'' \emph{arXiv preprint arXiv:1804.03641}, 2018.

\bibitem{kobar}
B.~Korbar, D.~Tran, and L.~Torresani, ``Cooperative learning of audio and video
  models from self-supervised synchronization,'' in \emph{Advances in Neural
  Information Processing Systems}, 2018, pp. 7763--7774.

\bibitem{speech2face}
T.-H. Oh, T.~Dekel, C.~Kim, I.~Mosseri, W.~T. Freeman, M.~Rubinstein, and
  W.~Matusik, ``Speech2face: Learning the face behind a voice,'' in
  \emph{Proceedings of the IEEE Conference on Computer Vision and Pattern
  Recognition}, 2019, pp. 7539--7548.

\bibitem{rochestergen}
L.~Chen, S.~Srivastava, Z.~Duan, and C.~Xu, ``Deep cross-modal audio-visual
  generation,'' in \emph{Proceedings of the on Thematic Workshops of ACM
  Multimedia 2017}, 2017, pp. 349--357.

\bibitem{visual2sound}
Y.~Zhou, Z.~Wang, C.~Fang, T.~Bui, and T.~L. Berg, ``Visual to sound:
  Generating natural sound for videos in the wild,'' in \emph{Proceedings of
  the IEEE Conference on Computer Vision and Pattern Recognition}, 2018, pp.
  3550--3558.

\bibitem{audio2body}
E.~Shlizerman, L.~M. Dery, H.~Schoen, and I.~Kemelmacher-Shlizerman, ``Audio to
  body dynamics,'' \emph{CVPR, IEEE Computer Society Conference on Computer
  Vision and Pattern Recognition}, 2017.

\bibitem{conversationalgestures}
S.~Ginosar, A.~Bar, G.~Kohavi, C.~Chan, A.~Owens, and J.~Malik, ``Learning
  individual styles of conversational gesture,'' in \emph{Proceedings of the
  IEEE Conference on Computer Vision and Pattern Recognition}, 2019, pp.
  3497--3506.

\bibitem{Musices}
H.~Zhou, Z.~Liu, X.~Xu, P.~Luo, and X.~Wang, ``Vision-infused deep audio
  inpainting,'' in \emph{The IEEE International Conference on Computer Vision
  (ICCV)}, October 2019.

\bibitem{music2gesture}
C.~Gan, D.~Huang, H.~Zhao, J.~B. Tenenbaum, and A.~Torralba, ``Music gesture
  for visual sound separation,'' in \emph{Proceedings of the IEEE/CVF
  Conference on Computer Vision and Pattern Recognition}, 2020, pp.
  10\,478--10\,487.

\bibitem{openpose}
Z.~{Cao}, G.~{Hidalgo Martinez}, T.~{Simon}, S.~{Wei}, and Y.~A. {Sheikh},
  ``Openpose: Realtime multi-person 2d pose estimation using part affinity
  fields,'' \emph{IEEE Transactions on Pattern Analysis and Machine
  Intelligence}, 2019.

\bibitem{multihead}
C.~S.~J. Doire and O.~Okubadejo, ``Interleaved multitask learning for audio
  source separation with independent databases,'' \emph{ArXiv}, vol.
  abs/1908.05182, 2019.

\bibitem{DResnet}
F.~Yu, V.~Koltun, and T.~Funkhouser, ``Dilated residual networks,'' in
  \emph{Computer Vision and Pattern Recognition (CVPR)}, 2017.

\bibitem{unet}
A.~Jansson, E.~Humphrey, N.~Montecchio, R.~Bittner, A.~Kumar, and T.~Weyde,
  ``{Singing voice separation with deep U-Net convolutional networks},'' in
  \emph{18th International Society for Music Information Retrieval Conference},
  2017, pp. 23--27.

\bibitem{ronneberger2015u}
O.~Ronneberger, P.~Fischer, and T.~Brox, ``U-net: Convolutional networks for
  biomedical image segmentation,'' in \emph{International Conference on Medical
  image computing and computer-assisted intervention}.\hskip 1em plus 0.5em
  minus 0.4em\relax Springer, 2015, pp. 234--241.

\bibitem{unet-img-synt}
G.~{Liu}, J.~{Si}, Y.~{Hu}, and S.~{Li}, ``Photographic image synthesis with
  improved u-net,'' in \emph{2018 Tenth International Conference on Advanced
  Computational Intelligence (ICACI)}, March 2018, pp. 402--407.

\bibitem{mao2016image}
X.~Mao, C.~Shen, and Y.-B. Yang, ``Image restoration using very deep
  convolutional encoder-decoder networks with symmetric skip connections,'' in
  \emph{Advances in neural information processing systems}, 2016, pp.
  2802--2810.

\bibitem{pix2pix}
P.~Isola, J.-Y. Zhu, T.~Zhou, and A.~A. Efros, ``Image-to-image translation
  with conditional adversarial networks,'' \emph{arxiv}, 2016.

\bibitem{Adam}
D.~P. Kingma and J.~Ba, ``Adam: A method for stochastic optimization,''
  \emph{CoRR}, vol. abs/1412.6980, 2014.

\bibitem{STFT}
``Chapter 7 - frequency domain processing,'' in \emph{Digital Signal Processing
  System Design (Second Edition)}, second edition~ed., N.~Kehtarnavaz,
  Ed.\hskip 1em plus 0.5em minus 0.4em\relax Burlington: Academic Press, 2008,
  pp. 175 -- 196.

\bibitem{sdr}
E.~{Vincent}, R.~{Gribonval}, and C.~{Fevotte}, ``Performance measurement in
  blind audio source separation,'' \emph{IEEE Transactions on Audio, Speech,
  and Language Processing}, vol.~14, no.~4, pp. 1462--1469, 2006.

\end{thebibliography}

\end{document}